# Sub-wavelength image manipulating through compensated anisotropic metamaterial prisms

**Junming Zhao, Yijun Feng\*, Bo Zhu, Tian Jiang**

*Department of Electronic Science and Engineering, Nanjing University, Nanjing, 210093, CHINA*
*\*Corresponding author: yjfeng@nju.edu.cn*

**Abstract:** Based on the concept of sub-wavelength imaging through compensated bilayer of anisotropic metamaterials (AMMs), which is an expansion of the perfect lens configuration, we propose two dimensional prism pair structures of compensated AMMs that are capable of manipulating two dimensional sub-wavelength images. We demonstrate that through properly designed symmetric and asymmetric compensated prism pair structures planar image rotation with arbitrary angle, lateral image shift, as well as image magnification could be achieved with sub-wavelength resolution. Both theoretical analysis and full wave electromagnetic simulations have been employed to verify the properties of the proposed prism structures. Utilizing the proposed AMM prisms, flat optical image of objects with sub-wavelength features can be projected and magnified to wavelength scale allowing for further optical processing of the image by conventional optics.

## 1. Introduction

One of the major obstacles in optics and photonics is the diffractive nature of light that has limited us to manipulate images at scales less than the wavelength of light [1,2]. Recently, the emerging field of metamaterials has provided ways to design artificial materials with unusual optical properties and such diffraction limit for imaging has been overcome through the perfect lens structure proposed by J. Pendry [3]. A planar slab of lossless left-handed metamaterial (LHM) with simultaneously negative permittivity and permeability can be made as a perfect lens, where both propagating and evanescent waves emitted from a given light source could be recovered completely at the image points inside and outside the slab. Although for practical realization, its sensitivity to material loss and other factors can limit the sub-wavelength resolution [4-7], the concept was validated experimentally by using a thin slab of structured LHM [8] at microwave range, or silver [9-10] and SiC [11] at optical range.

In addition to isotropic LHM, anisotropic metamaterials (AMMs) having permittivity and permeability tensors with parts of the elements being negative have drawn a lot of attentions due to their extraordinary properties, such as negative refraction [12,13], partial focusing [14], and resonance-cone focusing [15,16]. The AMMs have been identified into four types based on their wave propagation properties, which are called cutoff, always-cutoff, never-cutoff and anti-cutoff media [12]. Compensated bilayer structure of such AMMs has been proposed as an expanded perfect lens configuration, which could transfer the field distribution from one side of the bilayer to the other with sub-wavelength resolution not restricted by the diffraction limit [17,18]. Both theoretical analysis [19] and experimental verification [20] show that although such compensated bilayer lens provides no free space working distance, it could produce image with an enhanced resolution that exhibit a decreased sensitivity to losses and to deviations in material parameters relative to the LHM perfect lens configuration.

However, it has been pointed out that the perfect lens structure can only work for the near-field [21], which makes the image difficult to be processed or brought to focus by conventional optics. To solve this problem, a new sub-diffraction-limit imaging method called 'hyper-lens imaging' has been proposed [22-24] and experimentally verified [25-26], which can resolve and magnify sub-wavelength details utilizing the unusual optical phenomenon of strongly anisotropic metamaterials. The hyper-lens can project the magnified image into the far field - where it can be further manipulated by the conventional (diffraction-limited) optics. It should be mentioned that due to the cylindrical structure, such hyper-lens can only transfer image between the inner and the outer circular cylinder boundaries, which limits its optical applications. Further improvements have been reported by using the concept of coordinate transformation to design planar magnifying perfect-lens [27] or hyper-lens [28], but these theoretical proposals require complicated metamaterial with spatial varying anisotropic material parameters that are difficult to realize.

In this paper, we expand the concept of sub-wavelength image through compensated bilayer of AMMs. We propose two dimensional (2D) prism pair configurations of compensated AMM bilayer that are capable of manipulating sub-wavelength images. We demonstrate that image rotation with arbitrary angle, lateral image shift and image magnification could be achieved with sub-wavelength resolution through properly designed prism structures. Both theoretical analysis and full-wave electromagnetic (EM) simulations have been employed to verify the properties of the proposed prism structures. Utilizing the compensated AMM prisms, planar optical image of sub-wavelength objects can be magnified to wavelength scale allowing for further optical processing of the image by conventional optics.

## 2. Sub-wavelength imaging using 2D compensated bilayer of AMMs

Pendry's perfect lens can be viewed as a bilayer of vacuum and LHM, in which the LHM exactly compensates for the propagation effects associated with an equal length of vacuum. This concept of compensated bilayer has been extended to different types of AMM pairs [12,18], and 2D perfect lens structures could be accomplished by using AMMs for which some principal components of the permittivity and permeability tensors have negative value. For example, compensated bilayer of anti-cutoff (ACM) or never-cutoff medium (NCM) [12,19] could act as a perfect lens that transfers the electromagnetic field at the front surface of the bilayer to the back surface completely in both magnitude and phase to accomplish sub-wavelength resolution.

To understand the mechanism, we theoretically analyze the EM propagation when a 2D point source is placed outside a bilayer of AMM as shown in Fig 1. Assuming Region 0 and Region III are free space, while Region I and Region II are filled by AMMs with material tensor denoted as

$$\hat{\varepsilon}_j = \varepsilon_0(\varepsilon_{jx}\hat{x}\hat{x} + \varepsilon_{jy}\hat{y}\hat{y} + \varepsilon_{jz}\hat{z}\hat{z}), \quad \hat{\mu}_j = \mu_0(\mu_{jx}\hat{x}\hat{x} + \mu_{jy}\hat{y}\hat{y} + \mu_{jz}\hat{z}\hat{z}). \ (j = 1, 2). \quad (1)$$

To study the sub-wavelength imaging phenomenon, we assume the source is in front of the front interface of the bilayer with $d_1 \to 0$. Following the classical electromagnetic theory, the fields radiated by the point source can be expressed as closed forms of Sommerfeld-type integrals [29].

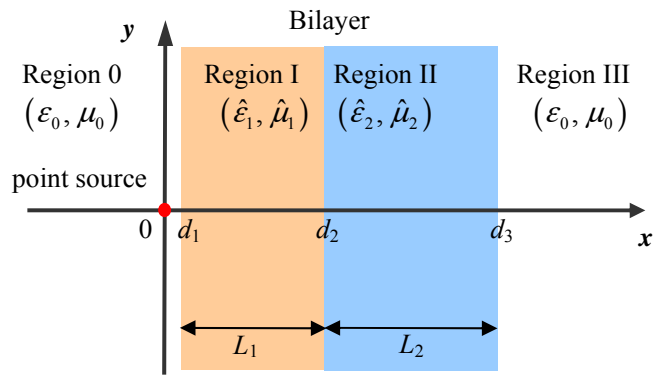

Fig.1. 2D point source propagating through a compensated AMM bilayer.

The transmission coefficient $T$ of the electric field at the back interface of the bilayer can be calculated as [29]

$$T = 8[(1+p)(1+q)(1+r)e^{-i(k_{1x}L_1+k_{2x}L_2)}+(1+p)(1-q)(1-r)e^{-i(k_{1x}L_1-k_{2x}L_2)} \\ +(1-p)(1-q)(1+r)e^{i(k_{1x}L_1-k_{2x}L_2)}+(1-p)(1+q)(1-r)e^{i(k_{1x}L_1+k_{2x}L_2)}]^{-1}, \quad (2)$$

where $p$, $q$, $r$ are the relative effective impedances defined as

$$p = k_{1x}/\mu_{1y}k_{0x}, \quad q = \mu_{1y}k_{2x}/\mu_{2y}k_{1x}, \quad r = \mu_{2y}k_{0x}/k_{2x}, \quad (3)$$

and $k_{0x}$, $k_{1x}$ or $k_{2x}$ are the $x$ component of the wave vector in free space, the first or the second layer of the bilayer, respectively, which satisfy different dispersion relations

$$k_y^2 + k_{0x}^2 = k_0^2, \quad k_y^2 + k_{jx}^2 \mu_{jx}/\mu_{jy} = k_0^2 \varepsilon_{jz} \mu_{jx}, \quad (j = 0, 1) \quad (4)$$

To obtain a perfect image at the back side of the bilayer, a unit transmission coefficient, $T = 1$, is required for all values of the transverse wave vector $k_y$. From Eq. (2), when $q = 1$, and $k_{1x}L_1 + k_{2x}L_2 = 0$ ($k_{1x}$ and $k_{2x}$ have opposite sign, representing propagating mode), or $q = -1$, and $k_{1x}L_1 - k_{2x}L_2 = 0$ ($k_{1x}$ and $k_{2x}$ have the same sign, representing evanescent mode), $T$ becomes unity resulting in a compensated bilayer [12].

Obviously, in addition of building compensated bilayer with vacuum and anti-vacuum ($\varepsilon = -\varepsilon_0$ and $\mu = -\mu_0$), which yields the configuration of Pendry's perfect lens, we could also build compensated bilayer by combining positive and negative refracting layers of NCM or ACM. The electromagnetic field at the front surface of the bilayer could be restored completely in both magnitude and phase at the back surface accomplishing near-field focusing. But we should emphasis the difference between isotropic Pendry's perfect lens and the AMM bilayer perfect lens. For example, in the case of NCM bilayer lens, both the propagating and the evanescent components of the source are converted into propagating modes in the NCM bilayer, and then back to propagating and evanescent components on the back surface, building an image with resolution beyond the diffraction limit.

The above analysis also allows us to construct compensated bilayer with unequal layer thickness. Assuming the thickness ratio $\eta = L_2/L_1$, compensated bilayer of NCM or ACM for both TE and TM waves requires

$$\hat{\varepsilon}_1 = \hat{\mu}_1 = \begin{bmatrix} \alpha_1 & 0 & 0 \\ 0 & -\beta_1 & 0 \\ 0 & 0 & \gamma/\alpha_1 \end{bmatrix}, \hat{\varepsilon}_2 = \hat{\mu}_2 = \begin{bmatrix} \alpha_2 & 0 & 0 \\ 0 & -\alpha_2\beta_1/\eta^2\alpha_1 & 0 \\ 0 & 0 & \gamma/\alpha_2 \end{bmatrix}, \quad (5)$$

where $\alpha_1\alpha_2 < 0$, and $\alpha_1$, $\alpha_2$, $\beta_1$, $\gamma$ are four arbitrary parameters that determine the material tensor elements of the compensated bilayer. In Eq. (5), $\gamma > 0$ stands for an ACM compensated bilayer, while $\gamma < 0$ stands for a NCM compensated bilayer.

We rigorously calculate the EM wave propagation of point sources through AMM compensated bilayer and two examples are demonstrated in Fig. 2. The first example is an equal thickness NCM bilayer with material parameters chosen as $\mu_{1x} = -\mu_{1y} = -1$, $\mu_{2x} = -\mu_{2y} = 1$, and $\varepsilon_{1z} = -\varepsilon_{2z} = 1$, and the second example is an unequal thickness NCM bilayer with parameters chosen as $\mu_{1x} = -\mu_{1y} = 1$, $\mu_{2x} = -3$, $\mu_{2y} = 1/3$, $\varepsilon_{1z} = -1$, and $\varepsilon_{2z} = 1/3$. In both case small electric and magnetic loss tangent of $10^{-3}$ are included to see their influences on the image quality. Excited with the two point sources that are separated about half wavelength at the front surface, two clearly focused images are observed at the back surface of the compensated bilayer as shown in Fig 2 (a) and (c). Unlike the case of an isotropic LHM planar lens, both the propagating and the evanescent components of the source incident into the NCM bilayer are converted into propagating modes, and then back to propagating and evanescent components on the back surface, building an image with resolution beyond the diffraction limit as shown in Fig. 2 (b) and (d). A standing wave mode is established inside the bilayer instead of a coupled surface plasmon mode in the case of LHM planar lens.

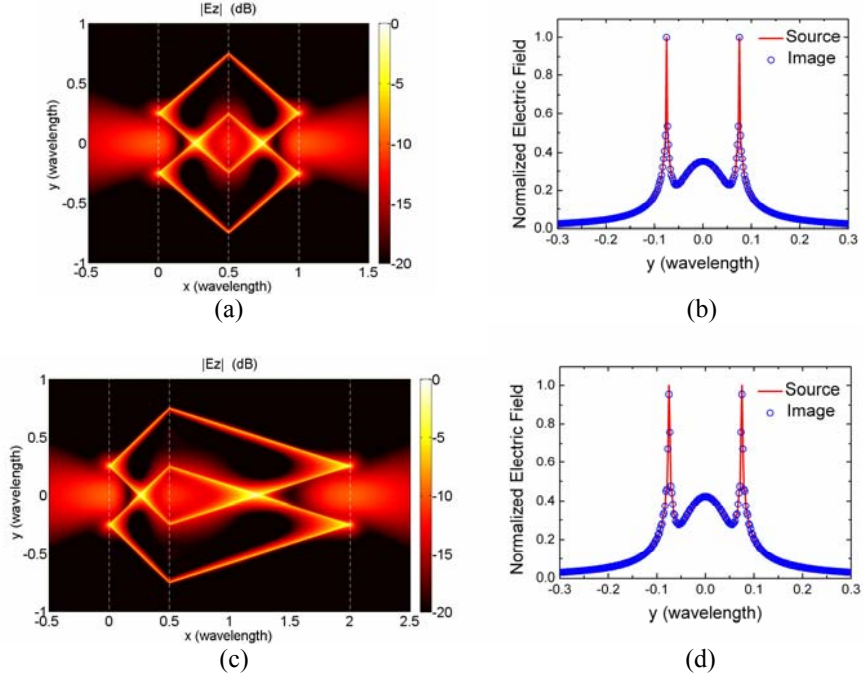

Fig. 2. Electric field mapping for two point sources located at the front interface of a NCM compensated bilayer with a loss tangent of $10^{-3}$ ((a), and (c)) and the comparison of the beamwidth of electric field distribution at the front and back interfaces ((b), and (d)) for an equal thickness bilayer ((a), and (b) with $L_2 = L_1$) and an unequal thickness bilayer ((c) and (d) with $L_2 = 3L_1$), respectively. The white dashed lines indicate the boundaries of the bilayer.

## 3. Sub-wavelength image manipulating through compensated AMM prism pairs

### 3.1 Sub-wavelength imaging between unparallel planes

The Pendry's perfect lens configuration is restricted to parallel source and image planes. Now we expand the compensated bilayer structure to a compensated prism pair (CPP) structure which can manipulate 2D images with sub-wavelength resolution between unparallel planes.

We first consider a symmetry CPP (S-CPP) configuration as shown in Fig. 3 (a). Two prisms with a same apex angle of $\theta$ (Prism1 and Prism2) are symmetrically putting together forming a prism pair. The two prisms are composed of AMMs with one of their optical axes (the $y$ axis) aligned with the symmetry axis OO′ and the material permittivity and permeability tensors satisfy the requirement of forming an equal thickness compensated AMM bilayer (i.e. Eq. (5) with $\eta = 1$). Due to the compensation nature of the two prisms, 2D sub-wavelength objects at the left surface OA will be perfectly imaged at the right surface OB of the S-CPP structure. For example, point sources at $S_1$ and $S_2$ of the source plane will be restored at $I_1$ and $I_2$ of the image plane, respectively, with $S_1I_1$ and $S_2I_2$ perpendicular to the interface OO′ of the structure. Such an S-CPP structure acts as an optical image component that makes perfect image between unparallel source and image planes with $2\theta$ rotation in the limit of sub-wavelength resolution.

To verify the performance, we carry out full wave EM simulation based on finite element method of the proposed S-CPP structure. Fig. 3 (b) shows the calculated electric field distribution of two point sources separated about 0.03 wavelength at the left surface of a S-CPP structure with $\theta = 30°$. The material tensor elements are chosen as $\mu_{1x} = -\mu_{2x} = -3$, $\mu_{1y} = -\mu_{2y} = 1/3$, and $\varepsilon_{1z} = -\varepsilon_{2z} = 1/3$, and a loss tangent of about 0.01 has been included for each tensor element. It is clearly demonstrated that at the right surface well resolved images of the two point sources are obtained, which confirms the ability of the S-CPP structure for imaging sub-wavelength objects to an unparallel plane. It is worth noting that the sub-wavelength imaging by S-CPP is not sensitive to small material losses, similar to that of the case of a compensated bilayer lens [19, 20]. Moreover, the boundary effect on the image quality is almost unobservable for prisms with finite sizes since most of the EM power is restricted to the rhombus area between the source and the image as indicated in Fig. 3 (b).

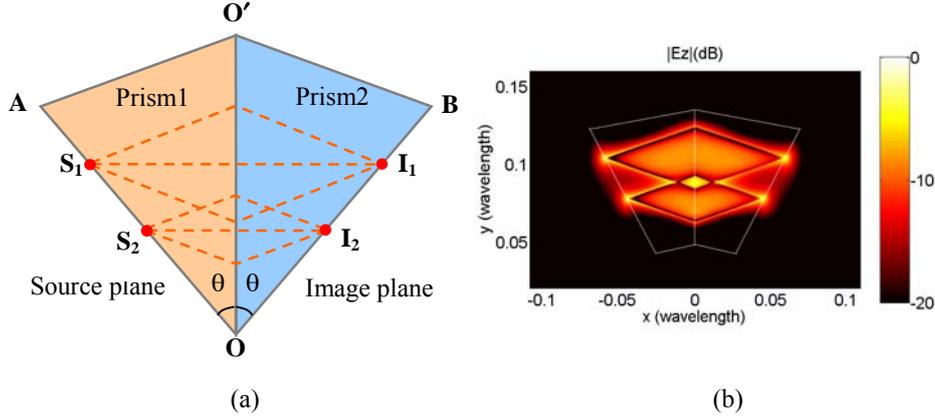

Fig. 3. (a) Schematic of a symmetry compensated prism pair configuration. (b) Electric field distribution of two point sources imaged with the S-CPP structure with a loss tangent of 0.01.

The proposed S-CPP structures can be used as optical components to build more complicated imaging system. For example, it can be cascaded to make different image rotation with arbitrary angles, or to produce sub-wavelength image with a lateral translation. Fig. 4 shows that a shifted image with sub-wavelength resolution can be obtained with a lateral translation to the original object by a combination of four identical S-CPPs with apex angle of $2\theta = 45°$. Two point sources separated by 0.04 wavelength have been well imaged at the other side of the S-CPP system with a 0.15 wavelength lateral shift.

*3.2 Magnified imaging with sub-wavelength resolution*

Next we consider more general case of an asymmetry CPP (AS-CPP) configuration as shown in Fig. 5. Two prisms with apex angles of $\alpha$ (for Prism1) and $\beta$ (for Prism2) are putting together forming an asymmetry prism pair. The two prisms are composed of AMMs with one of their optical axes (the $y$ axis) aligned with the common axis OO′ and the material permittivity and permeability tensors of the AMM satisfy the requirement of forming an unequal thickness compensated AMM bilayer (i.e. Eq. (5) with $\eta \neq 1$). Similar to the S-CPP structures, due to the compensation nature of the two prisms, sub-wavelength objects at the left surface OA will be perfectly imaged at the right surface OB of the AS-CPP structure. For example, point sources at $S_1$ and $S_2$ of the source plane will be restored at $I_1$ and $I_2$ of the image plane, respectively, with $S_1I_1$ and $S_2I_2$ perpendicular to the interface OO′ of the structure. The interesting feature of this AS-CPP structure is that the image size is unequal to that of the object with a magnification determined by

$$\tau = \frac{I_1 I_2}{S_1 S_2} = \frac{\cos \alpha}{\cos \beta}, \qquad (6)$$

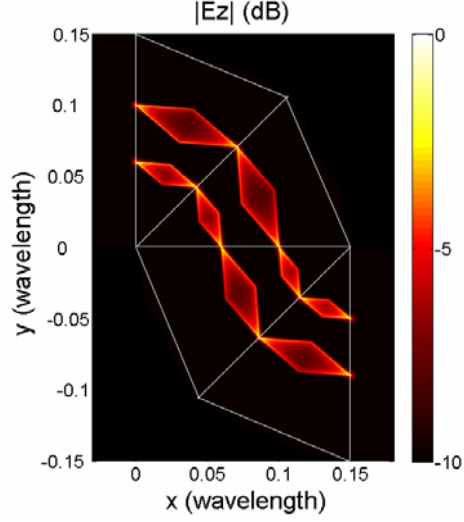

Fig. 4. Electric field distribution of two point sources imaged with four identical S-CPP structures cascaded together with a loss tangent of 0.01. Such configuration could produce lateral image translation with sub-wavelenth resolution. The white lines indicate the boundaries of the S-CPPs.

and the two apex angles $\alpha$ and $\beta$ are restricted by the compensation requirement of the two AMMs for the prisms, that is the thickness ration $\eta$ in Eq. (5), which satisfies

$$\eta = \frac{O_1 I_1}{S_1 O_1} = \frac{O_2 I_2}{S_2 O_2} = \frac{\tan\beta}{\tan\alpha}, \tag{7}$$

Thus, from Eq. (5) – (7) we are able to design certain AS-CPP structure that is possible of producing magnified image of sub-wavelength objects with in principle arbitrary magnification.

As an example, we design an AS-CPP with a magnification of $\tau = 3$. For convenient we can further restrict $\alpha + \beta = \pi/2$, then we yield $\eta = \tau^2 = 9$. The two apex angles of the prisms can be determined by Eq. (6) or (7), and the material parameters can be designed through Eq. (5). Actually we can see that there is plenty of freedom for design such an AS-CPP. Here we choose $\varepsilon_{1x} = \mu_{1x} = -3$, $\varepsilon_{1y} = \mu_{1y} = 1/3$, $\varepsilon_{1z} = \mu_{1z} = 1/3$, $\varepsilon_{2x} = \mu_{2x} = 27$, $\varepsilon_{2y} = \mu_{2y} = -1/27$, $\varepsilon_{2z} = \mu_{2z} = -1/27$, and a loss tangent of about 0.01 has been included for each tensor element. To verify the performance, we calculated the electric field distribution (Fig. 6 (a)) for three 2D point sources imaged with such an AS-CPP which has been cut into a cuboid shape. The material optical axes of the two prisms are parallel or perpendicular to the interface. The three point sources placed at the left boundary with unequal separations have been projected to the bottom boundary achieving a magnified image with sub-wavelength features of the objects. The line scans at both the source and image planes have also been compared in Fig. 6 (b) and (c), which clearly indicated a linear magnification of 3. The slightly intensity dimming and peak broadening from left to right of the image plane is due to the increasing loss as a result of the extending of the image-object distance, but the sub-wavelength features of the object have been well resolved.

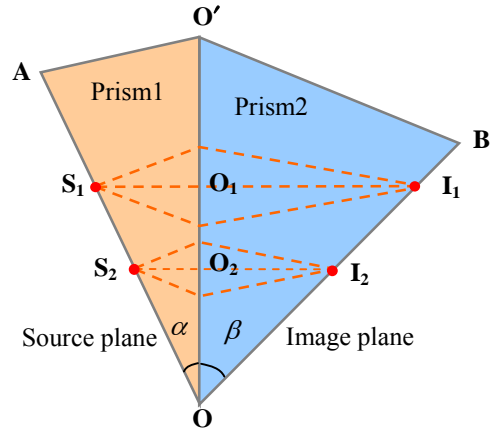

Fig. 5. Schematic of an asymmetry compensated prism pair configuration.

By cascading the AS-CPPs, we can make sub-wavelength image with larger magnification. Fig. 7 (a) shows the imaging of three point sources with two AS-CPPs cascaded together. The materials of the two AS-CPPs are similar to that used in the previous example, which lead to a total magnification of 9. The simulated electric field distribution is shown in Fig. 7 (a), and line scans at the source and image planes are illustrated in Fig. 7 (b) and (c), respectively, which indicate that the three point sources with separations within 0.1 wavelength have been well resolved and magnified to wavelength scale. The cascaded system also partially compensated the nonuniform image intensity in single AS-CPP resulting in more homogenous image intensity as shown in Fig. 7 (c). Using the AS-CPP structure, planar objects with deep sub-wavelength features can be projected and magnified to wavelength scale planar image. Such magnified image can be further processed by conventional optics, and the both flat object and image planes are more convenient for imaging and lithography applications.

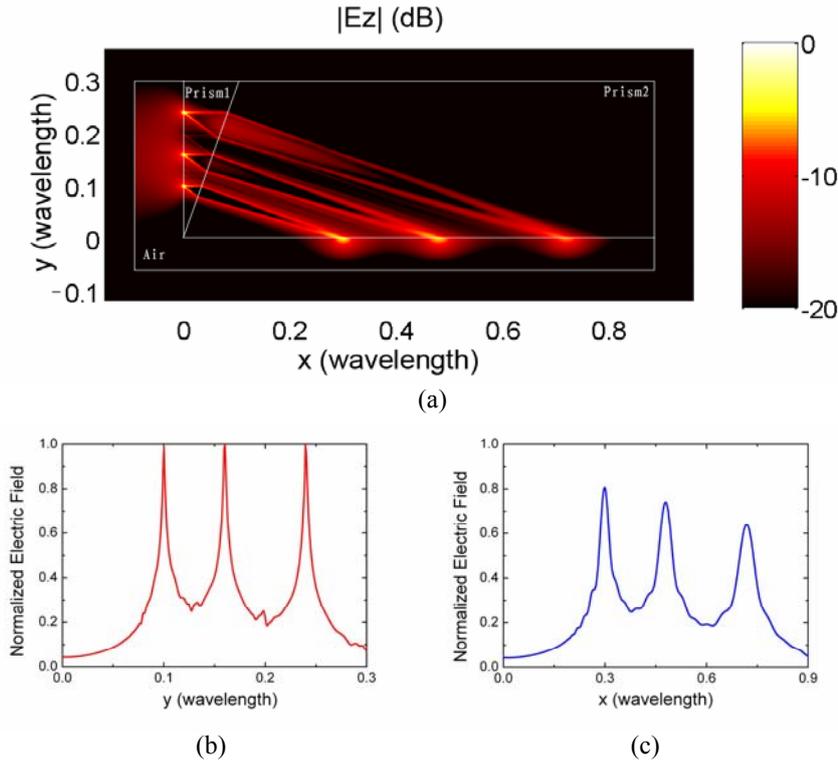

(a)

(b)  (c)

Fig. 6. (a) Electric field distribution for three point sources imaged with an AS-CPP with a designed magnification of 3. Loss tangent of 0.01 is included for each material parameter. Line scans at the source (b), and image (c) planes of the electric field which have been normalized to that of the source value.

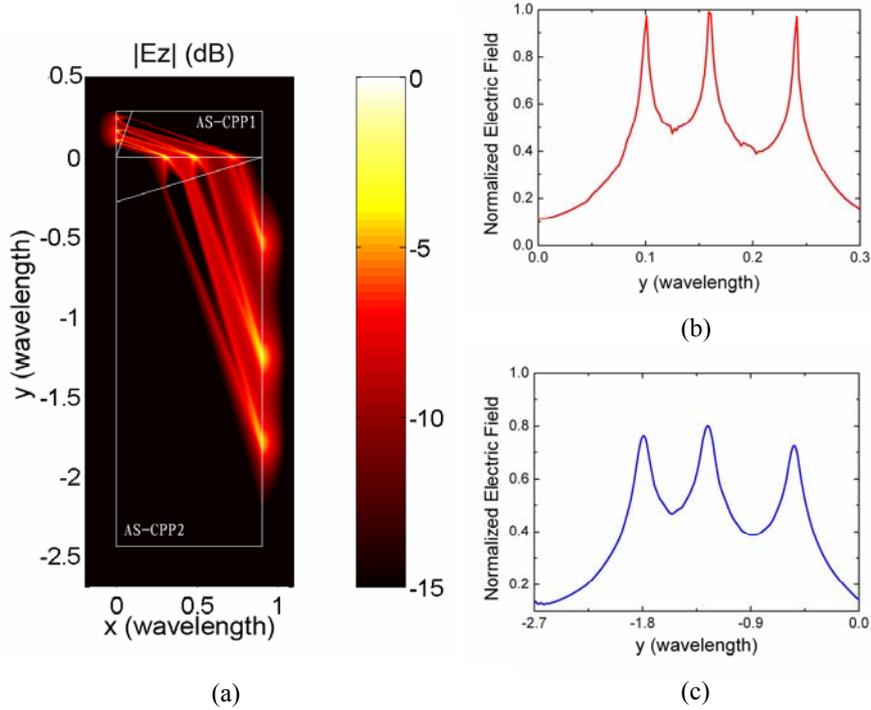

Fig. 7. (a) Electric field distribution for three point sources imaged with two cascaded AS-CPPs with a total designed magnification of 9. Loss tangent of 0.01 is included for each material parameter. Line scans at the source (b), and image (c) planes of the electric field which have been normalized to that of the source value.

The material requirements for building either A-CPP or AS-CPP are simply anisotropic with partial negative permittivity or permeability components and do not need any spatial variation. Thus they are more achievable compared to recent proposals of designing planar magnifying perfect-lens [27] or hyper-lens [28] based on the concept of coordinate transformation.

### 4. Conclusions

In this paper, we expand the concept of sub-wavelength imaging through compensated bilayer of AMMs. We propose 2D prism pair configurations of compensated AMM that are capable of manipulating sub-wavelength images. We demonstrate that planar image rotation with arbitrary angle, lateral image shift, as well as magnified image could be achieved with sub-wavelength resolution through properly designed compensated prism structures. The theoretical analysis and design procedure of these image processing components have been given and their performances have been confirmed by FEM based full wave EM simulation. Utilizing the proposed AMM prisms, planar optical image of objects with sub-wavelength features can be magnified to wavelength scale allowing for further optical processing of the image by conventional optics. With the rapid development of the metamaterial design and fabrication techniques in optical range, we believe the proposed sub-wavelength image manipulations could be applied to optical imaging and lithography systems with sub-wavelength resolutions.

**Acknowledgments**

This work is supported by the National Basic Research Program of China (2004CB719800), and the National Nature Science Foundation (No. 60671002).